\documentclass[aps,prl,twocolumn,showpacs,superscriptaddress,groupedaddress]{revtex4}  % for review and submission
\providecommand{\mbf}{\mathbf}

\usepackage{graphicx}  % needed for figures
\usepackage{dcolumn}   % needed for some tables
\usepackage{amsmath}   %for math
\usepackage{bm}        % for math
\usepackage{amssymb}   % for math
\usepackage{mymacros}

\usepackage{mathtools}
\usepackage{amssymb}
\usepackage{bm}
\usepackage{braket}

\RequirePackage{xspace}

\renewcommand*{\d}[1]{\mathrm{d}#1}

\renewcommand*{\vec}[1]{\mbf{#1}}

\begin{document}

\title{Kerr spacetime geometric optics for vortex beams
%General relativistic geometric optics for vortex beams in Kerr spacetimes
}

\author{Fabrizio Tamburini}
\email{fabrizio.tamburini@gmail.com}
\affiliation{ZKM -- Zentrum f\"ur Kunst und Medien,\\ Lorenzstra{\ss}e 19, D-76135 Karlsruhe, Germany}%
\author{Fabiano Feleppa}
\email{feleppa.fabiano@gmail.com}
\affiliation{Institute for Theoretical Physics, Utrecht University, Princetonplein 5, 3584 CC Utrecht, The Netherlands}%
\author{Ignazio Licata}
\email{ignazio.licata3@gmail.com}
\affiliation{Institute for Scientific Methodology (ISEM), I-90146 Palermo, Italy}%
\affiliation{School of Advanced International Studies on Theoretical and Nonlinear Methodologies of Physics, I-70124 Bari, Italy}
\affiliation{International Institute for Applicable Mathematics and Information Sciences (IIAMIS), B.M. Birla Science Centre, Adarsh Nagar, Hyderabad-500 463, India}
\author{Bo Thid\'e}
\email{bothide@gmail.com}
\affiliation{Swedish Institute of Space Physics, {\AA}ngstr\"om Laboratory, P.O. Box 537, SE-75121 Uppsala, Sweden}%

\begin{abstract}
We apply the analogy between gravitational fields and optical media in the general relativistic geometric optics framework to describe how light can acquire orbital angular momentum (OAM) when it traverses the gravitational field of a massive rotating compact object and the interplay between OAM and polarization.
Kerr spacetimes are known not only to impose a gravitational Faraday rotation on the polarization of a light beam, but also to set a characteristic fingerprint in the orbital angular momentum distribution of the radiation passing nearby a rotating black hole (BH). 
Kerr spacetime behaves like an inhomogeneous and anisotropic medium, in which light can acquire orbital angular momentum and spin-to-orbital angular momentum conversion can occur, acting as a polarization and phase changing medium for the gravitationally lensed light, as confirmed by the data analysis of M87* black hole.
\end{abstract}

\maketitle

%%% INTRO ASTRO

\section{\label{sec:level1}Introduction}
\vspace{-0.1cm}
Electromagnetic (EM) radiation is still the main carrier of information exploited to understand our Universe, to communicate and share information encoded in the EM conserved quantities and written in the observables of the EM field. 
The classical observables of the EM field are related to the set of the ten conserved quantities concomitant with the $10$-dimensional Poincar\'e group $P_{10}(EM)$ of Noether invariants: $3$ components of the linear momentum vector, $3$ of the pseudovector total angular momentum, $3$ components of the vector center of energy and the scalar energy. 
EM radiation transports both energy and momentum. 
The linear momentum vector is connected with force action, while the pseudovector total angular momentum $\mathbf{J} = \mathbf{S}+\mathbf{L}$ is connected with torque action. 
The spin-like form $\mathbf{S}$ is related to the spin angular momentum (SAM), associated with helicity, and hence with the polarization of light. The second form, $\mathbf{L}$, is associated with the orbital phase profile of the beam, measured in the direction orthogonal to the propagation axis, and is also known as orbital angular momentum (OAM).  
The transfer of SAM from light to a mechanical body was first demonstrated experimentally by Beth in 1935 \cite{Beth:PR:1935,Beth:PR:1936}, while the corresponding transfer of OAM has been verified only
recently \cite{Garces-Chavez&al:JOA:2004,Babiker&al:PRL:1994,andersen:170406}.
When a beam of light is propagating in inhomogeneous and anisotropic media, the conserved quantity $\mathbf{J}$ can evolve in time, together with $\mathbf{S}$ and $\mathbf{L}$, with an interchange between SAM and OAM 
\cite{2011JOpt...13f4001M,quantum1010010}.
Only in the paraxial approximation, when the beam is propagating in vacuum or in a homogeneous medium, $\mathbf{S}$ and $\mathbf{L}$ can be handled as two separate variables.
These properties remain valid down to the single photon level: each individual photon carries a quantized amount of SAM, $S=\sigma\hbar$, where  $\sigma=\pm1$, and can additionally carry OAM, whose $z$ component is quantized as well, $L_z=m \hbar$, where $m=0,\pm1,\pm2,\ldots,\pm{}N$ \cite{Berestetskii&al:Inbook:1980,Calvo&al:PRA:2006,2008PhRvA..78e2116T} as confirmed experimentally
\cite{Mair&al:N:2001,ONeil&al:PRL:2002,Leach&al:PRL:2002,Leach&al:PRL:2004}.

Laguerre-Gaussian (LG) helical beams are classical examples of beams with axial symmetry \cite{PhysRevA.94.023802}, characterized by inclined wavefronts that carry a well-defined value $L_z= m \hbar$ of OAM per photon~\cite{Schmitz&al:OE:2006,Molina-Terriza&al:NPH:2007} and harbor an EM or optical vortex (OV) in their center of symmetry. 
OVs are singularities of the EM field where the phase is not defined and the field amplitude drops to zero because of interference, forming in the plane transverse to the beam's propagation direction a doughnut-shaped structure with a central dark core where the vortex is found.

OAM of light finds practical applications in many fields of research. To give a few examples, we cite nanotechnology \cite{Grier:N:2003,nanotech2}, quantum cryptography \cite{Vaziri&al:JOB:2002,5689105,Mirhosseini_2015} and telecommunication \cite{Gibson&al:OE:2004,Thide&al:PRL:2007,Torres&Torner:Book:2011,Tamburini&al:APL:2011,Tamburini&al:NJP:2012} that has a vast literature with real-world demonstrations of efficient information transfer in the radio \cite{spinello2015experimental,triple,someda,oldoni} and  optical \cite{huang2014100} domains.

%%%%%%%%% ASTRONOMY
Nowadays OAM is attracting more and more the attention of the astronomical community.
Already in 2003, Harwit \cite{Harwit:APJ:2003} suggested to better exploit the properties of EM waves in order to optimally extract the information there encoded, and posed a particular attention to OAM.
Only recently OAM has become an efficient tool for astronomy as the first applications essentially consisted in the use of the optical properties of OVs to build efficient optical coronagraphs that can go beyond the Rayleigh limit: OVs obtained by artificially imposing OAM to the light coming from celestial bodies can enable the detection and imaging of extrasolar planets
\cite{Swartzlander:OL:2001,Swartzlander:OL:2005,Foo&al:OL:2005,Lee&al:PRL:2006,Anzolin&al:AA:2008}, and improve the resolving power of any optical instrument by up to one order of magnitude \cite{Tamburini&al:PRL:2006,PhysRevA.79.033845}.

The first remarkable result that demonstrated a clear advantage in the use of the OAM naturally emitted by celestial bodies with respect to the standard techniques used in astronomy is the measure of the rotation of the supermassive black hole (BH) in the center of the galaxy Messier 87 yielding a rotation parameter $a=0.90\pm0.05$ with $\sim 95\%$ confidence level \cite{10.1093/mnrasl/slz176}. 
This result was obtained from the analysis of the data collected by the Event Horizon Telescope collaboration (EHT) \cite{EHT1,EHT2,EHT3,EHT4,EHT5,EHT6} with standard OAM techniques \cite{Torner&al:OE:2005}, confirming that light propagating near a rotating black hole (Kerr) becomes twisted \cite{Tamburini&al:NPH:2011}.
The OAM analysis of the EM waves lensed by the BH is now a valid alternative and complementary method of investigation to the study of the shape of the BH shadow.

Theoretical studies indicate that EM waves can acquire OAM when propagating in a structured astrophysical plasma \cite{refId0}, providing information on the spatial structure and turbulence of the plasma \cite{Elias:AA:2008,Thide&al:Incollection:2011};
in this scenario, photons acquire an effective Proca mass through the Anderson-Higgs mechanism~\cite{Anderson:PR:1963,PhysRevLett.13.508}, becoming Majorana quasiparticles \cite{Tamburini_2011,PhysRevA.103.033505}.
This idea was applied to relativistic astrophysics to estimate the level of turbulence inside homogeneously distributed ultra-bright gamma ray burst (GRB) progenitors \cite{PhysRevD.96.104003,grbt}.
OAM applied to the BH information paradox demonstrated, in agreement with the latest Hawking's studies on soft-hair implants that the event horizon actually acts as a hologram, with a holographic memory that preserves the information carried by OAM states \cite{hawking0,hawking1,Tamburini&al:ENT:2017,hawking2}.

In this work we describe the evolution of the SAM and OAM components of a light beam propagating in a Kerr spacetime through general relativistic geometric optics.
In certain conditions the two components $\mathbf{S}$ and $\mathbf{L}$ of the total angular momentum invariant $\mathbf{J}$, ``disentangle'' their initial superposition, modifying and transforming SAM and OAM into each others as occurs in an inhomogeneous anisotropic medium.

\section{SAM and OAM evolution in Kerr spacetime}

To describe the evolution of the angular momentum of a light beam propagating in the spacetime of a rotating BH \cite{Kerr:PRL:1963}, we use the geometric optics-analogy of general relativity \cite{PhysRev.118.1396,defelice,2019JPhCS1330a2002R}. Geometric optics in GR suggests that light, when traversing regions of spacetime, behaves as it were propagating inside an optical medium. Following this analogy, we can characterize the SAM component of the beam with the gravitational Faraday effect  \cite{Dehnen:IJTP:1973} and study the evolution of the OAM component of $\mathbf{J}$. Different scenarios of rotating BHs due e.g. to modifications of the gravitational potential or modified theories of gravity go beyond the purpose of the present work.

Theory and the results from numerical simulations suggest that spacetimes of rotating BHs have geometries that are analogous to inhomogeneous and anisotropic media, as these beams acquire OAM during their propagation \cite{Tamburini&al:NPH:2011,Yang&Casals:PRD:2014} and change their polarization \cite{Su&Mallett:APJ:1980}. The analogies with particular optical media have been studied in deep and this analogy was tested in laboratory experiments \cite{10.1117/12.2582367} and can be in principle measurable with quantum information techniques that go down to the quantum level \cite{RACOREAN2018254}. 

Light beams and, more in general, EM waves emitted from an accretion disk (AD) around the rotating BH are characterized by specific fingerprints in the OAM spectrum \cite{Torner&al:OE:2005} that depend on the source (the physical properties of the AD) and on the GR effects i.e. the spacetime dragging and gravitational lensing. 

For the sake of simplicity, let us focus on the Kerr solution. 
In Boyer-Lindquist coordinates the line element of the Kerr spacetime with coordinates $(t,r,\theta,\phi)$ is given by the following quadratic form (for $G=c=1$) \cite{Ishihara&al:PRD:1988}
\begin{align}
ds^2&=\frac{\rho^2}{\Delta}dr^2+\rho^2d\theta^2
+\frac{\sin^2 \theta}{\rho^2}[adt - (r^2+a^2) d\phi]^2 \nonumber\\
&\hspace{0.4cm}-\frac{\Delta}{\rho^2}(dt-a\sin^2\theta d\phi)^2, 
\end{align}
where the coefficients
\begin{align}
\rho^2&=r^2+a^2\cos^2 \theta,\\
\Delta&=r^2-2Mr+a^2
\end{align}
depend on the mass $M$ and on the angular momentum per unit mass $a\leq{}M$ of the KBH.
Because of spacetime dragging, the effects of the gravitational lensing caused by a Kerr BH consist of image deformation and rotation of the lensed object; these effects are characterized by a particular anamorphic transformation, including additional features due to wavefront warping and gravitational Berry phase \cite{Carini&al:PRD:1992,1992mtbh.book.....C,Beckwith&Done:MNRAS:2005,Feng&Lee:IJMPD:2001,Yang&Casals:PRD:2014}.

The propagation of the polarization vector (SAM) along a null geodesic shows that the gravitational lensing produces a gravitational Faraday rotation \cite{1973IJTP7467D} of the plane of polarization with an angle proportional to the mass and the line-of-sight component of the angular momentum of the BH \cite{PhysRevD.38.472};
in addition to that, the BH can impose helical modes to the light which depend only on the rotation parameter $a$ of the BH and not on its mass $M$ \cite{Tamburini&al:NPH:2011}. A deep study and analysis of the spatial distribution of these two quantities, SAM and OAM, allows a better characterization of the phenomenology occurring in the regions surrounding the BH.
Of course, the distribution of the polarization vector and OAM in the observer's plane depends also on the intrinsic properties of the source before being lensed by the BH.
%SAM and OAM provide a different type of information about the BH properties. 
The different evolution of the polarization state (SAM) and OAM, the latter depending on the spatial distribution of the phase of the EM wavefront, can be put in evidence with some examples. Indeed, different polarization transformations can occur in a gravitational field that have the same initial and final polarization states but involves geometrical optical phase differences that are described by the Pancharatnam-Berry phase \cite{Feng&Lee:IJMPD:2001} and instead induce OAM. This effect can be traced and mathematically characterized by the analysis of the EM field of the lensed light with a multipole expansion \cite{Mashhoon:PRD:1973}. This procedure allows to identify and characterize the OAM through its projection on a specific orthogonal polynomial basis like Laguerre-Gaussian polynomials \cite{Molina-Terriza:PRA:2008}.  
%

%%% SAM and OAM \subsection{SAM and OAM interplay}

To study the transfer of OAM with the interplay between SAM and OAM we consider first a linearly polarized electromagnetic radiation propagating in a curved spacetime and then a circularly polarized one. 
In the first case, this wave experiences a rotation of the polarization plane due to the gravitational Faraday rotation effect.
On the other hand, when circularly polarized, light passes through a Faraday rotator and it experiences a gravitationally induced phase shift $\Delta \phi_G$ that depends upon the direction of propagation through the rotator.  
In this way, clockwise and counterclockwise polarized beams experience different phase shifts.

If the source of the gravitational field is rotating, the spacetime behavior changes dramatically.  
In this case, the norm of the timelike Killing field is positive in a region, called ergosphere, that extends outside the black-hole (BH) horizon;
the asymptotic time translation Killing field $\xi^\alpha=(\partial/\partial t)^\alpha$ becomes spacelike and an observer in the ergosphere cannot remain stationary even if it is orbiting outside the BH; observers
in the ergosphere are forced to move in the rotation direction of the Kerr BH (KBH).  

\subsection{Evolution of SAM and OAM in Kerr spacetime}

OAM is obtained from the spatial phase distribution imprinted by the gravitational lensing \cite{Carini&al:PRD:1992,Tamburini&al:NPH:2011,Yang&Casals:PRD:2014,10.1093/mnrasl/slz176}.
OAM can be generated or modified also by the SAM-to-OAM interchange that can occur in inhomogeneous and anisotropic media \cite{2011JOpt...13f4001M,Torres&Torner:Book:2011}.

Consider a typical case of gravitational lens where the source (s) and the observer (o) are located far away from the rotating BH \cite{Bray:PRD:1986} and the light passes close to the BH itself.  
We denote the coordinates of the light source by $(r_s,\theta_s,\phi_s)$ and the observer's coordinates by $(r_o,\theta_o,\phi_o)$; the deflection of a light ray from $s$ to $o$ occurs in a plane that includes the KBH too. We then introduce two local orthogonal reference frames defined by the scattering plane and the propagation directions of the light beam from (s) to the (o), $\mathbf{k_s}$ and $\mathbf{k_o}$, respectively.

As it is well-known, Kerr geometry belongs to the class of Petrov type D spacetimes \cite{2000GReGr..32.1665P}. The Walker-Penrose conserved quantity (along null geodesics) \cite{Walker:1970un, 1992mtbh.book.....C}, is used to describe the evolution of the polarization vector. The conserved complex quantity is
\begin{equation}
K_{\mathrm{WP}}=(A+\mathrm{i}B)(r-\mathrm{i}a\cos\theta),
\end{equation}
where, in turn, $A$ and $B$ are defined by
\begin{equation}
\begin{aligned}
  A&=(k^t f^r - k^r f^t) + a \sin^2 \theta (k^r f^\phi - k^\phi f^r),\\
  B&=(r^2 +a^2)\sin \theta (k^\phi f^\theta- k^\theta f^\phi) \\
  &\hspace{0.4cm}-a \sin \theta (k^t f^\theta - k^\theta f^t).
\end{aligned}
\end{equation}
In the above formulae, $k^i$ is the four-vector tangent to the geodesic and $f^i$ the polarization four vector that can be re-defined as $\hat{f}^r\equiv{}f^r$, $\hat{f}^\theta\equiv{}rf^\theta$, and $\hat{f}^\phi\equiv{}r\sin\theta{}f^\phi$.
In the region far away from the KBH, the real $\Re$ and imaginary $\Im$ parts of the invariant can be approximated as follows:
\begin{equation}
\begin{aligned}
  \Re (K_{\mathrm{WP}}) &= r(1-a u)\hat{f}^r+\gamma\hat{f}^\theta-\beta\hat{f}f^\phi,\\
  \Im(K_{\mathrm{WP}}) &= -r(1-a u)\hat{f}^r - (\beta\hat{f}^\theta+\gamma\hat{f}^\phi)\frac{k^r}{|k^r|}, 
\end{aligned}
\end{equation}
respectively.
Here, $u$ is a constant of motion that relates the asymptotic behavior of $k^i$ near the position of the source or of the observer,
$k^\phi\rightarrow\
u/(r^2\sin^2\theta)$, while
$\beta=(\eta - u^2\cot^2 \theta + a^2 \cos^2 \theta)^{1/2} k^\theta/|k^\theta|$ contains another constant of the motion, $\eta$.
Finally, $\gamma = u \csc \theta - a \sin \theta$.  
While the radial term $\hat f^r$ is characterized by a constant of the motion, the quantity $r (1-au)$,  depends directly on the radial coordinate and the transformation of the other two spatial components of the polarization vector transform from the source $(s)$ to the observer $(o)$ obtained with a linear rotation matrix $R$ whose coefficients depend on the geometry:
\begin{equation}
\left(\begin{array}{c}\hat f^\theta  \\ \hat f^\phi \end{array}\right)_o=R\left(\begin{array}{c}\hat f^\theta  \\ \hat f^\phi \end{array}\right)_s.
\end{equation}
Following Ref. \cite{Ishihara&al:PRD:1988}, we describe the evolution of the polarization vector
$\mathbf{f}=f_\parallel\mathbf{n}+f_{\perp}\mathbf{h}$ in terms of the perpendicular and projected components on the plane by transforming the polarization vector, expressed in the $(r,\theta,\phi)$ coordinates
\begin{equation}
N=\left(\begin{array}{cc}h^\theta & h^\phi \\n^\theta & n^\phi\end{array}\right).
\end{equation}
The final transformation is then
\begin{equation}
\left(\begin{array}{c} f_\parallel  \\ f_\perp
\end{array}\right)_o=N_oRN_s^{-1}\left(\begin{array}{c} f_\parallel  \\
f_\perp \end{array}\right)_s,
\end{equation}
giving the rotation of the polarization vector.
If the KBH spin is not lying in the propagation plane of the light beam, the rotation matrix of the polarization vector is reduced to the simple form
\begin{equation}
N_oRN_s^{-1}=\left(
\begin{array}{cc}
\cos \xi & -\sin \xi \\
\sin \xi & \cos \xi
\end{array}
\right).
\label{matJ}
\end{equation}

Let us now consider the two reference frames associated to $(s)$ and $(o)$.  The evolution of the polarization vector is fully described in the plane orthogonal to the propagation axis of the photons, identified by the vector components $f_\perp$ and $f_\parallel$.  
For a left-circular polarized beam ($f_\perp=1$ and $f_\parallel=\mathrm{i}$)
of coherent light that traverses the Kerr geometry of the rotating
gravitational lens, we similarly find that the lensed light acquires a
phase factor related to the OAM
\begin{equation}
N_oRN_s^{-1}\left(\begin{array}{c} 1  \\  \mathrm{i} \end{array}\right)_s=e^{-i\xi}\left(\begin{array}{c} 1  \\ - \mathrm{i} \end{array}\right)_o.
\end{equation}
In analogy with SAM-to-OAM conversion process \cite{Marucci&al:PRL:2006}, the OAM acquired by a light beam due to lensing is in our case $q = \xi/2$, which clearly depends on the phase retardation caused by the gravitational lens geometry as occurs in a spiral phase plate \cite{Torres&Torner:Book:2011}. 

In the observer's rest frame, the optical axis orientation is specified by an angle $\alpha(r,\phi) \propto q\phi$, where $q$ finds a relationship to the OAM value acquired by the circularly polarized light beam $m$ through $m = 2 q$.
Similarly as in Eq. (\ref{matJ}),
the Jones matrix
\begin{equation}
N(-\alpha)
\left(
\begin{array}{cc}
1 & 0 \\
0 & -1
\end{array}
\right)
N(\alpha)
=
\left(
\begin{array}{cc}
\cos 2\alpha & \sin 2\alpha \\
\sin 2\alpha & -\cos 2\alpha
\end{array}
\right)
\end{equation}
describes the change of the polarization vector due to the presence of the medium centered at $r=0$.  A left-circular polarized EM wave $E_s=E_0\times[1,\mathrm{i}]^T$, emerging from the medium, becomes right-circularly polarized with an additional phase factor $\exp(\mathrm{i}m\phi)$.  The beam has thus been transformed into a helical beam carrying a quantity of OAM $m=2q$. 
This suggests that, while a linearly polarized light experiences gravitational Faraday rotation, a circularly polarized beam instead can exhibit a behavior analogous to that experienced in an inhomogeneous anisotropic medium where spin-to-orbital angular momentum conversion occurs, a phenomenon related to the Pancharatnam-Berry geometrical phase %of inhomogeneous transformation of the polarization vector 
\cite{Marucci&al:PRL:2006} in a q-plate.

The phase evolution of a photon in a Kerr spacetime can also be described by using the photon wavefunction approach \cite{2008PhRvA..78e2116T}; different geodesics impose different real and imaginary values of the phase retardation, contributing to the gravitational Faraday rotation for
linearly polarized light and for the light intensity amplification.
The real phase factor depends on the coupling of the rotation of the BH with the photon SAM \cite{Carini&al:PRD:1992}. Furthermore, the phase evolution of photons can be also described in terms of Wigner rotation, tetradic representations of the curved spacetime \cite{Alsing&Stephenson:arXiv:2009}.

Thus, each photon has a finite and known probability of being converted from left-circular to right-circular, inducing an asymmetry in the helicity that depends on the mass and angular momentum of the rotating body, an effect related to the Pancharatnam-Berry geometrical phase for the evolution of the polarization vector. 
This asymmetry, which exhibits the gravitationally-induced spin transfer from the rotating body to the scattered photons, is the key of the spin-to-OAM conversion in a Kerr metric as no birefringence is present in GR \cite{Mohanty&Prasanna:arXiv:1995}.

With the same assumptions made in Ref. \cite{Bray:PRD:1986,Ishihara&al:PRD:1988}, including terms up to the third order in $M /r_{\mathrm{min}}$ and $a/r_{\mathrm{min}}$, we obtain the approximated solution to the equations of motion of a light beam that includes the multi-imaging aspect of the gravitational lens and the variation of the polarization.
In the small-deviations regime obtained for large values of $r$, the spatial coordinates used to describe the Kerr geometry can be approximated by a set of spherical polar coordinates $(r, \theta, \phi)$ in an Euclidean geometric support.
For each different light beam that draws the lensed image, the parameters $\nu$ and $\eta$ of the equations of motion can be expressed in terms of the so-called ``celestial coordinates'', denoted by $(\alpha,\beta)$, where $\alpha$ is the apparent perpendicular distance of the image from the axis of symmetry and $\beta$ the apparent perpendicular distance of the image from its projection on the equatorial plane \cite{1992mtbh.book.....C}. 
For the $i$-th light ray of the image, the constants of motion become
\begin{equation}
\nu \approx \frac{-r_0 \sin \theta_0 \alpha_i}{\sqrt{\alpha^2_i+\beta^2_i + (r_0 -\beta_i \cot \theta_0)^2}},
\end{equation}
and
\begin{equation}
\sqrt{\eta} \approx \frac{\mp r_0 \sqrt{\beta_i^2+\alpha^2_i \cos^2 \theta_0}}{\sqrt{\alpha^2_i+\beta^2_i + (r_0 -\beta_i \cot \theta_0)^2}}.
\end{equation}
The sign $\mp$ depends on the sign of the derivative $\frac{d \theta}{d r}|_{r=r_0}$ and this approximation excludes first- and higher-  order terms in $M r_{\mathrm{min}}^2/r_0^3$.
The expression for the distance of the closest approach $r_{\mathrm{min}}$ can then be written as a function of the celestial coordinates.
\begin{align}
r_{\mathrm{min}} &\approx \sqrt{\nu^2+\eta}\left[1- \frac M{\sqrt{\nu^2+\eta}}-\frac{3M^2}{2(\nu^2+\eta)}\right.
\nonumber \\
&\left. \hspace{0.4cm}-\frac {a^2}{2(\nu^2+\eta)}+\frac{a^2\eta}{2(\nu^2+\eta)^2}+ \frac{2aM \nu}{(\nu^2+\eta)^{3/2}}\right].
\end{align}
By converting the celestial coordinates of the $i$th light beam $(\alpha_i,\beta_i)$ in polar coordinates in the sky $(l,\varphi)$ centered on the KBH position, for the i-th beam, $(l_i,\varphi)$. Here $\varphi$ is the angle between the vector position of the $i$th image with the $\alpha$ axis, the real part of the phase factor acquired by photons $\xi$, expanded at the first order in $\varphi$, becomes
\begin{equation} 
\label{oambh}
\xi(\varphi) \simeq \frac {5 \pi}4 M^{2} a \cos \theta_0  \left[C_1 + C_2 \varphi\right] + O[\varphi]^2,
\end{equation}
where $C_1$ and $C_2$ are constants that depend on the parameters of the KBH and on the position of the observer,
%equation}%
\begin{align}
C_1 &= \left[\frac{a^2 r_0^2}2 -M + 
\frac{l_i^2 \sin^2 \theta_0 \sqrt{l_i^2+r_0^2}}{l_i^3 r_0^3} + \frac{l_i r_0}{\sqrt{l_i^2+r_0^2}} \right.
\nonumber\\
&\left. \hspace{0.4cm}-\frac{(a^2+3 M^2+2 a m \sin \theta_0 ) \sqrt{l_i^2+r_0^2}}{l_i r_0} \right]^{-3},
\end{align}
and 
\begin{align}
C_2 &= 3 \left[ 
\frac{(a^2+3M^2) \cot \theta_0 + 2a M \cos \theta_0 }{\sqrt{l^2_i+r^2_0}}
\right.
\nonumber \\
&\left. \hspace{0.4cm}+\frac{l^2_i r^2_0 \cot \theta_0}{(l_i^2+r_0^2)^{3/2}}
- \frac{l^2_i \sin \theta_0 \cos \theta_0}{l^2_i r^2_0 \sqrt{l^2_i+r^2_0}} 
\right] C_1^{-4/3},
\end{align}
where $C_1 = O[l]^0$ and $C_2=O[l]^{-1}$, respectively. Higher order terms in $\varphi$ have constants $C_j$ that decrease more rapidly with increasing $m$, being $O[C_j] \ll O[l]^{-2}$; thus, they can be neglected for our purposes.
The emission from the neighborhoods of any BH is limited by the photon sphere and by the size of the shadow \cite{feng}. With simple algebra, if one rescales the distances with respect to the photon orbit radius $r_g= 3 M$, it is easy to find that the OAM content of the lensed light depends on the rotation parameter $a$ only and not on the mass of the BH as numerically and experimentally proved in Ref \cite{Tamburini&al:NPH:2011,10.1093/mnrasl/slz176}.

In the weak gravitational regime the fraction of photons that experienced the spin-to-OAM conversion is fairly small. The first term, $C_1$, has the same order of magnitude as the classical gravitational Faraday rotation, while the term that describes the OAM component, is $C_2=O\left[10^{-11}\right]$.

In the framework of the analogy with geometric optics in GR, a good weak lensing approximation useful to estimate the content of OAM induced by the rotation of the KBH can be obtained from the geometrical considerations of the simplified Kerr lensing where the KBH clearly behaves as a holographic $q$ plate that imparts OAM onto the impinging light,
\begin{equation}
\label{lens1}
\langle m \rangle  \simeq \frac{5\pi}{4}\frac{M^2 a\cos\theta_0}{r^3_{\mathrm{min}}},
\end{equation}
where $r_{\mathrm{min}}$ is the shortest distance of the lensed null geodesic from the BH and $\theta_0$ is the projection angle on the line of sight of the component of the angular momentum of the KBH.
The reason we indicate the OAM content in Eq. (\ref{lens1}) with $\langle m \rangle$ is because OAM has a discrete spectrum of values. A non-integer value of OAM obtained by the gravitational lens is given by the superposition of the discrete OAM components present in the spiral spectrum.
As a rule of thumb, from the known results in Ref. \cite{Tamburini&al:NPHY:2011,10.1093/mnrasl/slz176},
the OAM content, namely the components different from the untwisted light ($m=0$) can be at the first order given by the ratio of the $m=1$ component with that of $m = 0$, viz.,
\begin{equation}
\langle m \rangle \simeq \mathrm{height}(m=1)/\mathrm{height}(m=0).
\end{equation}
The same order of magnitude is obtained with the gravitomagnetic approach
\cite{Guadagnini:PLB:2002,Barbieri&Guadagnini:NPB:2004}.

\begin{figure}
\includegraphics[width=9.6cm]{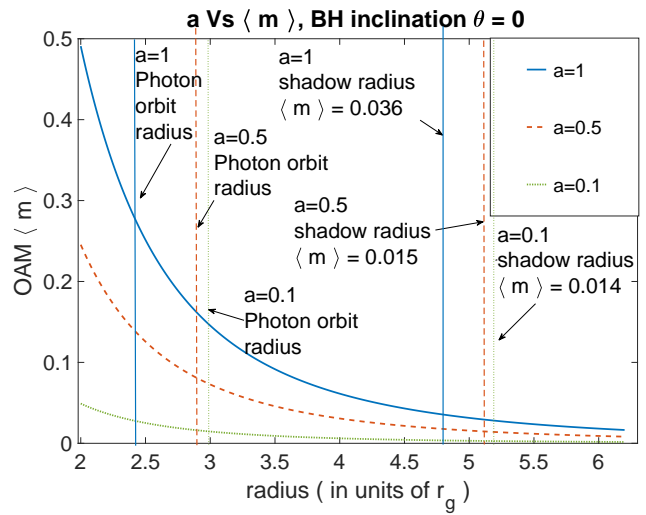}
\caption{Plot of the OAM value $\langle m \rangle$ of the radiation emitted by a particle falling into a KBH along the direction of the rotation axis versus the distance $2<r<6.2$ in units of $r_g$. The BH spin is pointing in the direction of the observer ($\theta=0$).
The quantity $\langle m \rangle$ is given by the ratio between the $m=1$ and the $m=0$ components in the spiral spectrum. Here are reported the values for three different angular momenta of the Kerr black hole: $a=0.1$ with $\langle m \rangle = 0.014$ (dot-dashed), $a=0.5$ with $\langle m \rangle = 0.015$ (dashed) and the extremal black hole, $a=1$ with $\langle m \rangle = 0.036$ (continuous plot).}
\label{Fig.1}
\end{figure}

Fig. \ref{Fig.1} shows the OAM value $\langle m \rangle$ acquired by a circularly polarized light beam lensed by three different KBHs with the rotation axis pointing to the observer's direction ($\theta_0=0$) \cite{feng}  having $a=0.1$, $a=0.5$ and $a=1$, respectively and finding  $\langle m \rangle_{a=0.1} = 0.014$ ,  $\langle m \rangle_{a=0.5} = 0.015$  and  $\langle m \rangle_{a=1} = 0.036$.
In the abscissa is reported the distance from the event horizon expressed in natural units ($c=G=1$) and the radius in terms of the photon orbit radius $r_g=3M$. The starting point of the x coordinate is at $r=2$, the ending point at $r=6.2$. 
The vertical asymptote (not displayed in the figure) corresponds to $r=0$ where the singularity is found and the OAM value acquired by light would tend to infinity. This is what could be observed with a naked singularity.

We now verify this model with the OAM content observed in the twisted light observed in M87*; 
let us assume as rotation parameter $a=0.90 \pm 0.05$ and inclination $i=17^\circ \pm 2^\circ$ (or equivalently $i=163^\circ \pm 2^\circ$ with a magnetic arrested disk) \cite{Tamburini&al:NPHY:2011}. The shadow of M87* is $\sim 2.6$ times larger than the black hole's Schwarzschild radius \cite{EHT1}. 
By using Eq. (\ref{lens1}), we find that the averaged OAM value $\langle m \rangle=0.0683$ observed in the experimental data from EHT from the analysis with the TIE method for epoch~1 (days April $5$\textendash April $6$) corresponds to a radius $r \simeq 3.5$. The value in epoch 2 corresponds to $r \simeq 3.4$ \cite{10.1093/mnrasl/slz176}, confirming that the OAM radiation is originated outside the BH shadow with null geodesics traveling in the neighborhoods of the BH. The result agrees with that obtained by computing the phase factor of Eq. (\ref{oambh}).

\subsection{OAM from image rotation}

%Rotation %%% image rotation
We now demonstrate, with another approach, the presence of OAM from the known phenomenon of image rotation induced by the Kerr geometry. 
When the source is in very close proximity of the BH, like in accretion disks, the value of OAM is quite significant and, for some particular cases, $m$ becomes larger than unity.
This Ansatz is based on the concept of equivalence of the curved spacetime to a medium with a precise index of refraction that affects the propagation of a light beam \cite{1980Ap&SS..68..221E}. 
The effect is analogous to the case in which the image rotation is obtained after the pencil of light has traversed a rotating medium.  More generally, rotating bodies and/or rotating observers experience a rotational coupling of the total angular momentum that splits into their extrinsic orbital and intrinsic spin parts. 
In the case of electromagnetic radiation, the orbital coupling is known in its limiting Doppler form, namely the Sagnac effect.

%When traversing the region of distorted spacetime, each photon has a certain probability of being converted from left-circular to right-circular, changing the component of the angular momentum from $+\hbar$ to $-\hbar$ and depends on the OAM acquired by the beam. 
The general phenomenon of spin-rotation coupling for photons has been the subject of investigations \cite{Courtial&al:PRL:1998}.
The OAM image rotation effect can be explained as being due to the phase shifts between the different constituent orbital angular momentum components of the image. 
The azimuthal phase term $\exp(-\mathrm{i}m\theta)$, which describes the helical phase fronts associated with orbital angular momentum (OAM), indicates that these modes have an $m$-fold rotational symmetry about the beam axis. We can describe this effect in analogy to the case in which the image rotation is obtained through a rotating medium: a rotation of the mode about this axis through an angle $\Delta \theta$ changes the phase by $\Delta \phi = m \Delta \theta$ and, although the image rotation is independent of $m$, the associated phase shift will be proportional to $m$.
The rotation of the plane of linear polarization is equivalent to a phase shift $\Delta \phi = (n_g - 1 / n_\phi) \Omega L$, where $\Omega$ is the angular velocity of the medium, like a glass rod, $n_g$ and $n_\phi$ are the refractive indices corresponding to the group and phase velocities of the light in the medium
and $L$ the thickness of the ``rod'' \cite{Padgett&al:OL:2006,Leach&al:PRL:2008}.

As the Laguerre-Gaussian modes form a complete basis set, any arbitrary image can be formed by an appropriately weighted superposition of these modes. The rotation of the image observed when the light is traversing a rotating medium is related to the OAM acquired by the photons \cite{Padgett&al:OL:2006,Leach&al:PRL:2008}. 
Similar results could be obtained when the light is transversing the distorted geometry in the close proximity of a Kerr black hole. In addition to the distortion caused by the gravitational lensing, calculations show that the lensed image is also rotated by the KBH, and in particular situations even up to a factor of $\pi /3$, when the source is directly behind or in front of the BH \cite{Pineault&Roeder:APJ:1977a,Pineault&Roeder:APJ:1977b}.

In our case, from simple geometrical considerations both for an accretion disk and for a lensed beam by a Kerr BH, the thickness of the ideal rod can vary in the interval $0<L<2r$. 
A reasonable approximation to estimate the order of magnitude of the OAM emitted is to assume $L \simeq r/2$.

In a particular geometrical formulation of the Kerr metric with cylindrical coordinates \cite{1980ApJ...238.1111S}, it was demonstrated that the light behaves like traversing an anisotropic medium. 
There, the direction of energy propagation given by the Poynting vector $\textbf{S}$ does not coincide with that of the wave vector $\textbf{k}$. This means that the group velocity $v_g$, associated to $\textbf{S}$, and the phase velocity $v_\phi$, associated with $\textbf{k}$ do not point in the same direction. The fields $\textbf{E}$, $\textbf{D}$ and $\textbf{H}$ are never collinear in this medium.
Letting $g_\alpha=- g_{0 \alpha}/g_{00}$ and $h=g_{00}$, the three-dimensional spatial metric tensor becomes $\gamma_{\alpha \beta}=-g_{\alpha \beta} + h g_\alpha g_\beta$ \cite{landau1975classical}. The fields can be written as $D_\alpha=E_\alpha/\sqrt{h}+g^\beta H_{\alpha \beta}$, $B^{\alpha \beta}= H^{\alpha \beta}/\sqrt{h}+g^\beta E^\alpha - g^\alpha E^\beta$, and we have also $B^\alpha=- 1/(2 \sqrt{\gamma})
e^{\alpha \beta \gamma} B_{\beta \gamma}$; the relations between the fields and the Poyinting and wave vectors are, as usual, $ \textbf{k}= \textbf{D} \times \textbf{B}$ and $\textbf{S}= \textbf{E} \times \textbf{H}$.

The analogy with a rotating glass rod, of course different for each null-geodesic, follows from the Hamilton-Jacobi formalism and Fermat's principle. 
The value of the refractive index $n$ for the Kerr metric, as a function of the rotation parameter $a$, obtained with a post-Newtonian approximation \cite{2003NCimB.118..249I}, is 
\begin{align}
\label{refind}
n &\simeq \frac 1{\sqrt{1-\frac{\tilde{r}}{\rho^2}}}\left[1+ \frac{5 a^2 r \sin^2 \theta_0}{\tilde{r} \left(\rho^2 - \tilde{r}\right) \left(1+\sqrt{1-\frac{4a^2}{r^2_g}}\right)^2} \right],
\end{align}
with $\tilde{r}=2M$. To obtain just an order of magnitude of the phase change $\Delta \phi$, we approximate the region nearby the rotating BH by a rod with the thickness of the order of the gravitational radius, rotating with the angular velocity of the KBH at a radius $r'$ (expressed in units of the gravitational radius), namely 
\begin{equation} \label{rotation}
\Delta \phi = a m \left(n - \frac1n \right) r',
\end{equation}
for an external observer  $r'> r_{sh}$, considering that the main contributions of OAM is generated in the regions close to the BH; in certain cases $r'$ can be even smaller than the radii of rotating and counterrotating photon circular orbits 
(see e.g. ref.  \cite{1972ApJ...178..347B}).

We can see, from the results obtained with M87*, that this very simple model of a ``rotating rod'' can give a good description of the Kerr spacetime as an optic medium, confirming the validity of the GR geometric optics approach also with twisted light. For M87*, from Eq. (\ref{refind}), the refraction index for $r=3.5$ is $n=1.16$ and at the shadow's radius, $r_{sh}=2.6$, one finds $n_{sh}=1.28$. From this one can easily calculate the ``thickness'' of the ideal rotating rod, finding $L \sim r/2$ and confirming our initial assumptions.

%If we consider as example an extremal KBH of $4 M_\odot$. The refraction index $n=1.24$ is obtained for $r'=1.3$ and a beam with OAM value $m\simeq1.8$, a superposition of $m=1$ and $m=2$ modes. 
%%
%With this approach higher OAM terms with $m > 2$ in the spiral spectrum do not provide useful additional information to the BH rotation, as confirmed by the observations of M87* and with the method in Ref. \cite{10.1093/mnrasl/slz176} to determine the BH rotation parameter $a$.

\section{conclusions}

In this work we discussed the evolution and conversion of the angular momentum of light beams and photons from the spin to the OAM component when they traverse the region of a rotating black hole. 
As in optical inhomogeneous and anisotropic media, Kerr lensing can generate helical modes of light that depend on the geometry and dynamics of the lensing body. 
Moreover, different polarization transformations with the same initial and final polarization states involve geometrical optical phase differences described by the Pancharatnam-Berry phases, a principle that can be extended to a more general form of wave fronts shaped by polarization transformations where OAM and optical vorticity are included.

With this simplified approach we found that the orbital angular momentum quantity $m$ acquired by a light beam lensed by a Kerr black hole can exceed unity, carrying a quantity $m \hbar$ of orbital angular momentum per each photon present in the beam.
This result finds agreement with the values of $m$ found with numerical simulations and confirmed experimentally with the analysis of the electromagnetic waves received from the neighborhood of the black hole in M87* with radio telescopes.

%Alternatively, with a more complete approach, as made in Ref. \cite{10.1093/mnrasl/slz176}, one can obtain OAM numerically from the the Stokes parameters $(I,U,V)$ for the linear and elliptic polarization states as reference fields, the Pancharatnam-Berry phase of the vortex field $\E(x,y)$ is given by the the ratio between the circular/elliptic state of polarization of the EM field, characterized by the Stokes parameter $V$, and the state
%${\left|\Phi_V\right\rangle}$, and that of the initial field $\E(x,y)$, namely, ${\left|\Phi_\E\right\rangle}$.

According to Ref. \cite{Zhang&al:SR:2015}, one obtains the OAM $m$ from the intensity, the gradient of the spiral spatial phase distribution (obtained from the Stokes parameters $V$ and $U$) and the variation of the state of the polarization in space at each point in the observational plane of the asymptotic observer (o).

For the sake of completeness only, one can in principle obtain higher values of Faraday rotation on the order of $\Delta \phi_G \sim \pi$ with the approach described in Ref. \cite{1980ApJ...238.1111S}, implying a relevant contribution of the components in the OAM spectrum with $m > 1$ following the analogy with the phase modifying device in Ref. \cite{Marucci&al:PRL:2006}. 
This occurs when the sources are located in the neighborhoods of the event horizon or when light rays, lensed from infinity, travel very close to the BH event horizon, a scenario expected when the BH is fast-rotating with $a\sim1$ or when the BH is a fast spinner. 
The results obtained so far suggest that a relevant presence of very high order components in the OAM spectrum could reveal the presence of naked singularities.

Of course, the evolution of SAM and OAM and SAM-to-OAM transfer is in principle valid also for other fields crossing the gravitational field of the KBH.  In particular, neutrino fields and gravitational waves may reveal the hidden latest evolution stages of a supernova or a GRB during its collapse when the forming KBH is setting its rotation, in an attempt to preserve the total angular momentum while interacting with the surrounding matter and fields. Helicity asymmetry and OAM can in fact be induced in both these fields \cite{Barbieri&Guadagnini:NPB:2005}.

\section*{Acknowledgements}The authors thank Dennis Durairaj for his suggestions.

%%%%
%We point out that in those considerations we assumed that all the light beams impinging the distorted geometry were beams of coherent light. Astrophysical sources of coherent light are quite common. LASER and MASER sources are present in the sky associated with certain stars or gas clouds, namely galactic or circumstellar masers (for a review see \cite{Elitzur:ARAA:1992}).
%Astronomical masers are often very highly polarized and the most common polarization state is circular, with some less common examples of linear polarized sources. Water masers are found in dense molecular clouds closely associated with supermassive black holes at the centres of active galaxies.
%
%Astrophysical scenarios described by the Kerr metric can be approximated
%in the low-energy domain by gravitoelectromagnetism, an approximation in
%which EinsteinÕs GR can be described with a mathematical
%structure similar to Maxwell's electrodynamics
%\cite{Mashhoon:Incollection:2006}.  

\end{document}